\documentclass[10pt, journal]{IEEEtran}

\usepackage{cite,graphicx,epsfig, psfrag, subfigure, url, amsmath, array}
\usepackage{algorithm, algorithmic, calc, color}
\usepackage{graphics}
\usepackage{epsfig}
\usepackage{amssymb}
\usepackage{multicol}

\newtheorem{example}{Example}

\renewcommand{\baselinestretch}{1}

\newcommand{\ls}[1]
   {\dimen0=\fontdimen6\the\font
    \lineskip=#1\dimen0
    \advance\lineskip.5\fontdimen5\the\font
    \advance\lineskip-\dimen0
    \lineskiplimit=.9\lineskip
    \baselineskip=\lineskip
    \advance\baselineskip\dimen0
    \normallineskip\lineskip
    \normallineskiplimit\lineskiplimit
    \normalbaselineskip\baselineskip
    \ignorespaces
   }

\begin{document}

\ls{1}

\title{Opportunistic Network Coding for Video \\Streaming over Wireless}

\author{\authorblockN{Hulya Seferoglu, Athina Markopoulou\\}
\authorblockA{Electrical Engineering and Computer Science Dept.\\
University of California, Irvine\\ Irvine, CA, USA\\
{\em \{hseferog,athina\}@uci.edu}}}

\maketitle

\begin{abstract}
In this paper, we study video streaming over wireless networks with
network coding capabilities. We build upon recent work, which
demonstrated that network coding can increase throughput over a
broadcast medium, by mixing packets from different flows into a
single packet, thus increasing the information content per
transmission. Our key insight is that, when the transmitted flows
are video streams, network codes should be selected so as to
maximize not only the network throughput but also the video quality.
We propose video-aware opportunistic network coding schemes that
take into account both (i) the decodability of network codes by
several receivers and (ii) the importance and deadlines of video
packets. Simulation results show that our schemes
significantly improve both video quality and throughput.
\end{abstract}

\begin{keywords} Video streaming, wireless networks, mesh networks, network coding.\end{keywords}

\section{Introduction}

Providing high quality video over wireless networks is a challenging
problem, due to both the erratic and time-varying nature of a
wireless channel and the stringent delivery requirements of media
traffic. Developments in video compression and streaming, wireless
networking, and cross-layer design, are continuously advancing the
state-of-the art in wireless video \cite{wvbook,wv-specialissue}. In
this paper, we propose a novel technique for video streaming in a
wireless environment inspired by the emerging paradigm of network
coding \cite{netcodingwebpage,FragouliCCR06}.

Our work builds on recent work in \cite{cope,xor} that used network
coding to improve throughput in a wireless mesh network. In
particular, \cite{cope,xor} proposed that wireless routers mix
packets from different flows, so as to increase the information
content of each -broadcast- transmission and therefore the
throughput for data applications. In this paper, we build on this
idea, and propose a network coding and scheduling scheme for
transmitting several video streams over a wireless mesh network.

Our key insight is that the transmission of video streams in a
network coding-capable wireless network should be optimized not only
for network throughput but also, and more importantly, for
video quality. The fact that video packets have unequal importance
is well understood and extensively studied in the video streaming
community, e.g. for rate-distortion optimized streaming \cite{radio,
radio-jacob, radio-mark}. The fact that mixing different information
flows can increase throughput in multicast networks is well
understood in the network coding community
\cite{netcodingwebpage,FragouliCCR06,nc1,nc2}. Our work bridges the
gap between the two approaches, and proposes a new video-aware
scheme for network coding and packet scheduling that improves both
aspects, namely video quality and throughput.

We consider a wireless mesh network, in which routers can mix
different incoming flows/streams, using simple network coding
operations (XOR). The resulting network code is broadcasted to the
neighborhood of the router. Nodes in the same neighborhood listen to
each other's transmission and store overheard packets; these are
used later to decode received coded packets and also to construct
new coded packets. The core question in this architecture is how to
select the best -according to an appropriate metric- network code
for transmission among all possible codes. In \cite{cope,xor}, a
transmitting node chooses a network code that can be decoded by
several neighbors  at the same time slot; this policy increases the
utility of each transmission thus leading to throughput benefits.
However, when the transmitted flows are video streams, this is not
necessarily the best choice. Video quality can be improved by
intelligently selecting network codes that combine those video
packets that are decodable by several neighbors but also contribute
the most to video quality. In other words, when video streams are
transmitted, it is not only the quantity but also the
quality/content of information transferred that should be taken into
account in the selection of network codes. In this paper, we develop
schemes for network code selection and packet scheduling that take
into account both (i) the importance and deadlines of video packets
and (ii) the network state and the received/overheard packets in the
neighborhood.

The paper is organized as follows. Section \ref{sec:related}
discusses related work. Section \ref{sec:system} gives an overview
of the system model. Section \ref{sec:algs} presents the algorithms
for network coding.
Section \ref{sec:performance} presents simulation results that demonstrate
the benefits of the proposed algorithms over baseline schemes, in
terms of video quality and application-level throughput. Section
\ref{sec:discussion} discusses open issues and ongoing work. Section
\ref{sec:conclusion} concludes the paper.

\section{\label{sec:related}Related Work}

This work combines ideas and techniques from two bodies of work:
video streaming and network coding.

Several network-adaptive techniques have been proposed to support
streaming media over unreliable and/or time-varying networks
\cite{networkadaptive}. Supporting video over wireless is
particularly challenging due to the limited, time-varying resources
of the wireless channel \cite{wvbook,wv-specialissue}. There is a
large body of work on cross-layer design for video over wireless,
including \cite{mihaela, crosslayer-2, crosslayer-3,
crosslayer-4,crosslayer-5}. Packet scheduling is an important
control at the medium access control layer. The problem of
rate-distortion optimized packet scheduling has been studied in the
RaDiO family of techniques \cite{radio, radio-mark, radio-jacob}: in
every transmission opportunity, media units are selected for
transmission so as to maximize the expected quality of received
video subject to a constraint in the transmission rate, and taking
into account transmission errors, delays and decoding dependencies.
Cross-layer approaches exploit the fact that packets in a video
stream have different importance and therefore should be treated
differently by network mechanisms.

Independently, the network coding paradigm has emerged from the
pioneering work in \cite{nc1, nc2}, which showed that, in multicast
networks where intermediate nodes do simple linear operations on
incoming packets, one can achieve the min-cut throughput of the
network to each receiver. The linearly combined packets can be
utilized at the receivers to recover the original packets by solving
a set of linear equations over a finite field. This breakthrough
idea inspired significant effort in several directions
\cite{netcodingwebpage, FragouliCCR06}, including studying
topologies beyond multicast, such as unicast
\cite{li-unicast,chou-unicast,ho-unicast} and broadcast scenarios.
The broadcast nature of the wireless medium
 offers an opportunity for
exploiting the throughput benefits of network coding
\cite{medard-wireless, nc-wireless}. The recent work in \cite{cope,
xor} applied these ideas from the network coding community in the
 context of wireless mesh networks. \cite{xor} implemented
a pseudo-broadcast mechanism for 802.11 together with opportunistic
listening and a coding layer between IP and MAC that is used to
detect coding opportunities and pack packets from different flows
into a single transmission, thus increasing network throughput.

Our paper introduces a novel technique for video streaming over
wireless that combines the above two approaches. On one hand, we
build on \cite{cope,xor} to exploit the broadcast nature of the
wireless medium and use network coding to pack several packets {\em
from different streams} into a single code for transmission, thus
increasing throughput. On the other hand, we construct and select
network codes taking into account the importance of video packets
(in terms of video distortion and playout deadlines) {\em within the
same stream}, as well as their contribution to the total throughput
and video quality. This combined approach allows us to achieve
significant video quality improvement while still maintaining the
throughput benefits.

\section{\label{sec:system}System Overview}

\begin{figure}
\centering \vspace{-30pt}
\includegraphics[width=8.2cm]{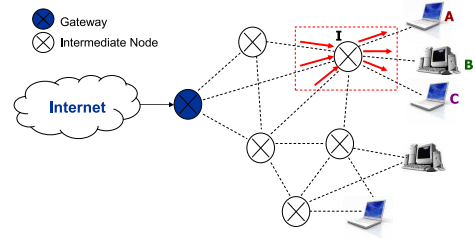}
\vspace{-40pt} 
\caption{A wireless mesh network} \label{fig1}
\vspace{-10pt}
\end{figure}

We consider video streaming over wireless mesh networks where
intermediate nodes (wireless mesh routers) are able to forward
packets to other intermediate nodes and/or clients, as shown in
Fig.~\ref{fig1}. In this paper, we propose algorithms that can be
used at the intermediate node to maximize video quality and
throughput. We assume that intermediate nodes can perform simple
network coding operations (bit-wise XOR) and combine packets from
several incoming streams into a single outgoing packet. This packet
is broadcasted to the entire neighborhood, thus reaching several
nodes at the same time. We assume that nodes can overhear all
transmissions in their neighborhood, whether they are intended for
them or not; they can decode a network-coded packet using overheard
packets. The idea of combining network coding with broadcast to
increase the information content per transmission, is well
understood in the network coding community. This idea has been
recently applied in 802.11-based multi-hop wireless networks and
throughput benefits have been demonstrated for data applications
\cite{cope,xor}.

Our key observation is that, when the transmitted flows are video
streams, this is not necessarily the best choice and video quality
must also be considered. The importance and deadlines of video
packets must be taken into account to intelligently select those
network codes that contribute the most to the quality of video
streams.  In this paper, we develop schemes for network coding
across different flows, and packet selection within each flow, to
improve both video quality and throughput.

{\em Code Selection at the Intermediate Node.} Let us consider an
intermediate node that receives $N$ packets from different video
streams and forwards them to $N$ nodes in its neighborhood. The
intermediate node maintains a {\em transmission (Tx) queue} with
incoming video packets. At a given time slot a packet is selected
from the Tx queue for transmission. The selected packet is called
the {\em primary packet} and its destination node is called the {\em
target node}. Depending on the network coding scheme, the primary
packet may be the first packet from the head of the queue, or any
packet in Tx queue that is marked as {\em active} ({\em i.e.} not
transmitted within the last round-trip time). In addition to the
primary packet, all packets in the queue are considered as candidate
{\em side packets}, i.e. candidates for a transmission in the same
time slot together with the primary packet. The primary and the side
packets are all XOR-ed together into a single packet, called the
{\em network code}.\footnote{The primary packet can be thought as
the main packet we try to transmit during a time slot; this is the
packet that would be normally transmitted by a FIFO policy. Side
packets are opportunistically transmitted together with the primary
packet; they are useful to nodes other than the target node.} The
core question then is:
\begin{quote}
{\em which network code (i.e. XOR of the primary and side packets)
to select and transmit so as to maximize the total video quality and
throughput.}
\end{quote}
The algorithms addressing this question are the main part of this
paper, and will be discussed separately in the next section
(\ref{sec:algs}). In the rest of this section, we describe the
remaining components and functions of the system. The terminology is
summarized in Table \ref{table:notation}.

\begin{table}[t]
\caption{\label{table:notation}Terminology} \vspace{-20pt}
\begin{center}
\begin{tabular}{|p{1.8cm}|p{5.6cm}|}
  \hline
Term & Definition\\
\hline Primary Packet & The packet selected from the Tx queue before
network coding. It must be included in all network codes.
It can be thought as the main packet we try to transmit in a given time-slot.\\
Side Packet & Packet in the Tx queue, other than the primary, included in the network code.\\
Active Packet & Packet in the Tx queue that can be considered as
primary. (Not transmitted within
the last RTT.)\\
Inactive packet & Packet in the Tx queue that cannot be considered
as primary. (It has already been transmitted within the last RTT,
and
the acknowledgement is still pending.) \\
Network Code & The primary and side packets XOR-ed together into a single packet.\\
Target Node & The intended recipient of the primary packet.\\
Tx Queue & The output queue of the transmitting node. 
\\
Rx Buffer & The receive queue of the receiving node. It stores received packets,
destined to this node.\\
Virtual Buffer & Also maintained at a receiving node. It stores overheard packets,
destined to other nodes.\\
  \hline
\end{tabular}
\end{center}
\vspace{-15pt}
\end{table}

{\em ACKs and Other Functions at the Receiving Nodes.} Once the
network code is chosen, it is broadcasted to all nodes in the
neighborhood. Depending on the channel conditions, some nodes
successfully receive it. When the target node receives it, it
decodes it (which is guaranteed by the construction of the network
code in the next section), stores the primary packet in its {\em
receive (Rx) buffer}, and sends an acknowledgement (ACK) back to the
intermediate node. Nodes, other than the target node, overhear the
transmitted packet and try to decode it; if they overhear a new
packet destined to them, they store it in their Rx buffer and send
an ACK back to the intermediate node; if they obtain a packet
destined for another node, they store it in their {\em virtual
buffer}. An overheard packet stays in the virtual buffer until an
ACK from the target is overheard or until its deadline expires.

{\em Retransmissions and Active/Inactive Packets at the Intermediate
Node.} The intermediate node waits for a mean round-trip time (RTT)
from the time it transmits the network code until it receives an
ACK. During that period, all packets that were part of the code stay
in the Tx queue but are marked as {\em inactive}. Inactive packets
are not considered for primary transmission (in order to avoid
unnecessary duplicate transmissions) but are still considered as
candidates for side packets (to increase coding opportunities). When
the transmitter receives an ACK, it removes the corresponding packet
from the Tx queue. If an RTT expires without receiving an ACK, the
packet is marked as {\em active} again and the process is repeated.
A packet stays in the Tx queue until either it is successfully
transmitted or its deadline expires; when either of these occur, the
packet is removed from the transmission buffer.

Notice that considering any active packet as primary, better
utilizes the bandwidth but may lead to reordering. Although this may
be a concern for TCP, it is clearly better for video that requires
timely delivery and can reorder packets at the playout buffer.

{\em Requirements.} We assume the following capabilities available
at our system. First, {\em broadcast} is needed to harvest the
benefits of network coding. Although wireless is inherently a
broadcast medium, this may be hidden by some communication
protocols. We assume that some broadcast capability is available,
e.g. 802.11 broadcast or pseudo-broadcast as implemented in
\cite{cope,xor}. Second, nodes need to know the {\em contents of the
virtual buffers} of all their neighbors, in order to code. In our
simulations we assume perfect knowledge of the contents of all
virtual buffers. This can be achieved by exchanging and guessing
this information, as in \cite{cope,xor}; in practice, there will be
some error and a slight degradation in the overall performance.
Third, nodes must be capable of {\em coding/decoding} in real time,
which is a realistic assumption for simple (bit-wise XOR)
operations. Finally, we assume a Tx queue with video packets only.

\section{\label{sec:algs}Coding Algorithms}

The main questions in this system have to do with the construction
and selection of network codes. The {\em code construction} problem
is concerned with finding candidate codes that guarantee
decodability by the target and several other nodes. The {\em code
selection} problem is concerned with finding the best among the
candidate codes to optimize video quality and throughput.
The first proposed algorithm, NCV, achieves the same throughput
gains as in \cite{xor} but also intelligently chooses the packets to
improve video quality. The second algorithm, NCVD, uses NCV as a
building block but considers more coding options, thus further
improving video quality and throughput.

\subsection{NCV Algorithm: Network Coding for Video}

Assume that there are several video streams coming to an
intermediate node that can be mixed together. Depending on the
content of virtual buffers at the clients, there may be several
combinations of these streams, i.e. several network coding
opportunities. The main idea behind the Network Coding for Video
(NCV) algorithm is to select the best network code to improve video
quality. Let us demonstrate this idea through an example.

\begin{figure}[t!]
\centering \vspace{-30pt}
\includegraphics[width=8.2cm]{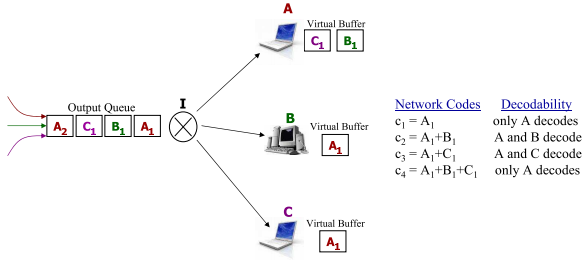}
\vspace{-50pt} \caption{Example of Network Coding for Video (NCV)}
\label{fig:ncv}
\end{figure}

\begin{example}\label{ex1}
Consider the example shown in Fig.~\ref{fig1} and let us focus on a
single-hop shown in more detail in Fig.~\ref{fig:ncv}. Node $I$
receives three independent video streams, e.g. from the Internet
through the gateway, destined to its neighbors $A,B,C$. $I$
maintains a FIFO Tx queue that stores packets $\{A_1,A_2,...\}$
destined to node $A$, $\{B_1,B_2,...\}$ destined to node $B$, and
$\{C_1,C_2,...\}$ destined to node $C$. Fig.~\ref{fig:ncv} also
shows the contents of the virtual buffers at each client: node $A$
has overheard packets $\{B_1,C_1\}$ and nodes $B$ and $C$ have both
overheard packet $A_1$, from previous transmissions. $A_1$ is the
first active packet from head of the queue and is selected as the
primary packet. Any packet (active or inactive) in the output queue,
other than $A_1$, can be chosen as a side packet,
on the condition that the constructed network code should be decoded
at node $A$, i.e. $A_1$ can be retrieved. To satisfy this condition,
side packets that will be used in the network code should already be
available at node $A$; in other words, the decodability of a network
code depends on the overheard packets at node $A$. Network codes
$c_1 = A_1$, $c_2 = A_1\bigoplus B_1$, $c_3 = A_1\bigoplus C_1$, and
$c_4 = A_1 \bigoplus B_1 \bigoplus C_1$ can all be decoded by $A$
and thus are eligible network codes.
$\Box$
\end{example}

{\em The Code Construction Problem.} More generally, consider an
intermediate node $I$ with $N$ clients $\{n_1, n_2, ..., n_{N}\}$.
There are $\Phi$ packets in the Tx queue, which are denoted with
$\{p_{1}, p_{2}, ..., p_{\Phi}\}$. Choose the first active packet,
$p_{i}$, from the head of the Tx FIFO queue as the primary packet
with target node $n_m$. $I$ will construct and transmit a network
code, which consists of $p_{i}$ XOR-ed together with some side
packets. The network code should be constructed so as to guarantee
decodability of the primary packet at its target node $n_m$. For
$p_{i}$ to be decodable at $n_m$, all $n-1$ side packets must be
among the overheard packets at $n_m$. Assume that $\Psi_m$ packets
are overheard at node $n_m$ and denoted by $\{\nu_{m,1}, \nu_{m,2},
..., \nu_{m,\Psi_m}\}$.
 Therefore, the candidate network codes are:
$$c_k^{i} = \{p_{i}\} \bigcup S_k^{m}, k=1,2, ...,
2^{\Psi_m}$$ where $S_k^{m}$ is the $k^{th}$ subset of $\{\nu_{m,1},
\nu_{m,2}, ..., \nu_{m,\Psi_m}\}$. Note that, since
linear operations are limited to bit-wise XOR, a network code $p_1\bigoplus p_2
\bigoplus {...} \bigoplus p_k$ is completely specified by the set of
packets $\{p_1,p_2,...,p_k\}$ that are XOR-ed together. The
complexity of considering all possible network codes is discussed in
section \ref{sec:complexity}. The next step, is to select the best
among all candidate codes.

{\em Example 1 Continued.} Node $A$ can get packet $A_1$ from all
possible network codes. Codes $c_2$ and $c_3$ improve the video
quality at node sets $\{A,B\}$ and $\{A,C\}$, respectively. It is
clear that $c_2$ and $c_3$ are better codes than $c_1$ and $c_4$
both for throughput (they are useful to two instead of one node) and
video quality. Comparing  $c_2$ to $c_3$, we observe that they are
equivalent in terms of throughput but they may contribute
differently to video quality depending on the content of video
packets $A_1, B_1, C_1$. Deciding which candidate code to select
between $c_2=A_1 \bigoplus B_1$ and $c_3=A_1 \bigoplus C_1$ should
depend on the importance and urgency of the original video packets
$B_1$ and $C_1$. NCV exploits this observation. $\Box$

\begin{algorithm}[t!]
 \caption{The NCV Algorithm} \label{algo:NCV}
\begin{algorithmic}[1]
\begin{footnotesize}
\STATE Initialization: $I_{max}^{i}=0$, $c_{max}^{i}=\emptyset$
\STATE Choose the first head-of-queue active packet as primary
$p_{i}$. \STATE Let $n_m$ be the target node of $p_i$. Let
$\{\nu_{m,1}, ..., \nu_{m,\Psi_m}\}$ be the overheard packets at
$n_m$.
\FOR {$k=1...2^{\Psi _m}$} \STATE $c_k^{i}=\{p_{i}\}\bigcup
S_k^{m}$ \STATE Calculate $I_k^{i}$ with Eq.~(\ref{eqNo1})
\IF{$I_k^{i}>I_{max}^{i}$} \STATE $I_{max}^{i} = I_k^{i}$,
$c_{max}^{i} = c_k^{i}$ \ENDIF \ENDFOR \STATE Choose
$c_{max}^{i}$ as the network code. XOR all packets and transmit
\end{footnotesize}
\end{algorithmic}
\end{algorithm}

{\em The Code Selection Problem.} In order to choose the best code,
we first need to define a metric that captures the contribution of
each candidate code to video quality improvement. Assume that
$p_{i}$ is the primary packet targeted to node $n_m$, and
$\{c_k^{i}\}_{k=1}^{k=2^{\Psi_m}}$ are all the candidate codes. Let
$I_k^{i}(n_m)$ be the improvement in video quality at client $n_m$,
when $c_k^{i}$ is received and decoded:
\begin{equation} \label{eqNo2} I_k^{i}(n_m) =
\sum_{l=1}^{L_k}(1-e_{l}^{k})\Delta_{l}^{k}g_{l}^{k}d_{l}^{k},
\end{equation}
where each factor in this formula is defined as follows:
\begin{itemize}
\item$L_k$ is the number of original packets included in network code
$c_k^{i}$. \footnote{Notice that at most one out of these $L_k$
packets can be useful to a particular node $n_m$, but different
packets are useful to different nodes.}
\item $d_{l}^{k}$ and $g_{l}^{k}$ are indicator functions that express
whether code $k$ is useful for node $n_m$. We define $d_{l}^{k}=1$
if $c_k^{i}$ is decodable at node $n_m$, or 0 otherwise. We define
$g_{l}^{k}=1$ if packet $l$ is targeted to node $n_m$, or $0$
otherwise.
\item $\Delta_{l}^{k}$ is the improvement in video quality (SNR) if
packet $l$ is received correctly and on time at client $n_m$. To
compute $\Delta_{l}^{k}$, we decode the entire video sequence with
this packet missing and we compute the resulting distortion.
\footnote{This is an approximation as the actual distortion that may also
depend on the delivery status of prior and subsequent NALs. The
distortion model can be extended to capture these loss correlations
\cite{burstlength}. Furthermore, we assume that distortions caused
by loss of multiple packets are additive, which is reasonable for
sparse losses. These approximations reduce the computational
complexity by separating the total distortion function into a set of
individual packet distortion functions and optimizing for each one
of them.}
\item $e_{l}^{k}$ is the loss probability of packet $l$ due to
channel errors or latency:
\begin{equation} \label{eqNo3}
e_{l}^{k}=\int _{\tau}^{\infty} p_F(t)dt + \left (1 - \int
_{\tau}^{\infty} p_F(t)dt\right)\varepsilon_F(s).
\end{equation}
The first part in Eq.(\ref{eqNo3}) describes the probability of a
packet arriving late; $\tau$ is the remaining time until the playout
deadline and $p_F(t)$ is the distribution of the forward-trip time.
The second part describes the loss probability (of a packet that is
still on time) due to effects of the wireless channel, such as
noise, fading, interference, etc; $\varepsilon_F(s)$ is the loss
probability at state $s$ of the channel.
\end{itemize}
After defining the contribution of code $c_k^{i}$ to a single node
$n_m$, $I_k^{i}(n_m)$, we define the total video quality improvement
of code $c_k^{i}$ as the sum of the video quality improvements at
all clients $n_1, ...n_N$,\footnote{If the quality of the encoded
video sequences to different clients $n_m$ are significantly
different
from each other,
then the terms $I_k^{i}(n_m)$ should be normalized by the average PSNR per sequence,
before adding them up.} due to code
$c_k^i$:
\begin{equation} \label{eqNo1}
I_k^{i} = \sum _{m=1}^{N} I_k^{i}(n_m),
\end{equation}

The NCV algorithm is summarized in Alg.~(\ref{algo:NCV}). At each
time slot, the NCV algorithm chooses the primary packet $p_{i}$ and
constructs all candidate network codes
${c_k^i}_{k=1}^{k=2^{\Psi_m}}$. Among all candidate network codes,
NCV chooses the code that maximizes the total video quality
improvement:
\begin{equation} \label{eqNo0}
\max_k I_k^{i}
\end{equation}
Notice that, depending on the contents of the virtual buffers, it is
possible that no side packets can be used together with a given
primary packet $p_i$. In that case, the network code is simply
$\{p_i\}\cup \emptyset =\{p_i\}$, and we transmit only the primary
packet alone.

\subsection{NCVD Algorithm: looking into the queue in Depth}

\begin{figure}[t!]
\centering \vspace{-30pt}
\includegraphics[width=8.2cm]{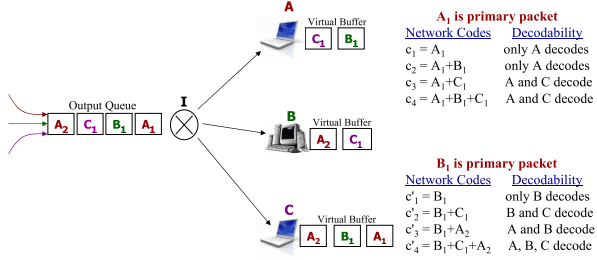}
\vspace{-50pt} \caption{Example of NCVD}\label{fig:ncvd}
\end{figure}

As described in the previous section, NCV selects the primary packet
from the head of the queue but ignoring packets marked as
``inactive''. This limits the candidate codes to those that are
decodable for this single primary packet. The second algorithm, NCVD
looks into the entire Tx queue (``in depth'') and considers all, not
just the head-of-line, packet as candidates for the primary packet,
thus increasing the options for candidate codes, which eventually
leads to a better choice for the metric of interest. Note that a
different set of candidate codes can be constructed for each primary
packet. Let us explain NCVD through the following example.

\begin{example}\label{ex2}
Let us look at Fig.~\ref{fig:ncvd}. The topology  is the same as in
Fig.~\ref{fig:ncv}, but the contents of the Tx queue and of the
virtual buffers are different. Assume that all packets are active
packets, i.e. they can all be considered as primary. One option is
to select the head-of-line packet $A_1$ as the primary packet. As
discussed in Example 1, the best codes for this primary packet are
$c_3=A_1\bigoplus C1$ or $c_4=A_1 \bigoplus B_1 \bigoplus C_1$. A
different choice is to select $B_1$ as the primary packet, which
leads to completely different set of candidate network codes (listed
on the Fig. \ref{fig:ncvd}). Code $c_4' = B_1 \bigoplus C_1
\bigoplus A_2$ achieves the maximum throughput improvement, and
potentially the maximum video quality achievement, depending on the
importance and urgency of all packets. This example demonstrates
that increasing our options of primary packet, increases the set of
candidate codes, and thus can potentially improve both throughput
and video quality. $\Box$
\end{example}

More generally, NCVD constructs candidate codes $c_k^{i},k=1,2, ...,
2^{\Psi_m}$ for each candidate primary packet $p_{i}$ in the Tx
queue. Among all constructed codes, NCVD selects the code that
maximizes the total improvement in video quality for all clients:
\begin{equation} \label{eqNo4}
\max _{p_{i}} \max_k (I_k^{i}),
\end{equation}
Algorithm \ref{algo:NCVD} summarizes NCVD.

NCVD can be parameterized by the depth $d$ of the Tx queue, that is
considered in the selection of the primary packet. NCVD($d=1$) is
simply NCV, while NCVD($d=\infty$) considers all packets in the Tx
queue. The larger the value of $d$, the more coding options, the
better the performance of NCVD. Because queue sizes are small for
real time applications, we can focus on NCVD($d=\infty$), which we
simply refer to as NCVD.

\begin{algorithm}[t!]
 \caption{The NCVD Algorithm}
\label{algo:NCVD}
\begin{algorithmic}[1]
\begin{footnotesize}
\STATE Initialization: $c_{max} = \emptyset$, $I_{max}=0$ \FOR{every
packet $i=1,..., ,\Phi$ from the head of Tx queue} \STATE Consider
this packet, $p_{i}$, as candidate for primary \STATE Construct
all possible codes $c_k^{i}$ for $p_{i}$\STATE
Determine the max improvement
$I_{max}^{i}=\max_{k} I_k^{i}$ \\
\STATE and the corresponding code $c_k^{i}$: $k=argmax I_k^{i}$
as in NCV \IF{$I_{max}^{i}>I_{max}$} \STATE $I_{max} =
I_{max}^{i}$, $c_{max} = c_k^{i}$ \ENDIF \ENDFOR \STATE Choose
$c_{max}$ as the network code. XOR all packets and transmit.
\end{footnotesize}
\end{algorithmic}
\end{algorithm}

\section{\label{sec:performance}Performance Evaluation}
In this section, we evaluate the performance of the proposed schemes
({\em NCV} and {\em NCVD}) in terms of video quality and network
throughput under different conditions. We compare them to two
baseline schemes, one without coding ({\em noNC}) and one with
network coding optimized for throughput ({\em NCT}) as in
\cite{xor}. Simulation results show that NCV and NCVD can
significantly improve video quality and application-level
throughout, without compromising MAC-level throughput.

\subsection{Simulation Setup}
Let us first describe the communication scenario, wireless channel
models and video sequences used, and the baselines algorithms for
comparison.

\subsubsection{Single-Hop Scenario}
In this paper, we focus on the evaluation of the single-hop scenario
shown in Fig.~\ref{fig:ncv}: the intermediate node $I$ receives
different video streams which it forwards downstream towards
their destinations $A,B,C$. $I$ can apply different schemes for
network coding and packet scheduling in the downlink. The downlink
rate is selected to be 300 kbps, and each video stream has a rate of
70 kbps. We assume that the three video streams are the only ones
using the downlink, hence, there is no congestion. However, packets
may still be lost due to error on the wireless channel, according to
models described in the next paragraph, and can also experience a
random delay, which we assume to be exponentially distributed with
$4ms$ average. The delay budget (playout deadline) for this single
hop is set to $100ms$, to allow for queueing, transmission,
propagation and a few retransmissions. We also perform simulations
for different delay budgets, from 50ms to 200ms. ACKs and
retransmissions are used to combat errors, as explained in the
system overview.

\subsubsection{Wireless Channel Model}
We consider two channel models to evaluate our algorithms in
different conditions. In both cases, packet loss is independent and
identically distributed across the three sessions (from $I$ to $A$,
from $I$ to $B$, and from $I$ to $C$). Below we specify the packet
loss model within a session.

{\bf Model I:} Packets transmitted in each link experience i.i.d.
loss.  with a fixed loss probability. We simulated a wide range of
effective packet loss rates from 1\% up to 20\%.
The effective loss rate depends on the use of  retransmissions, FEC
and other mechanisms that mask the error rate on the channel.

{\bf Model II:} A flat Rayleigh fading channel is modeled by a
finite-state Markov chain (MC), where the fading is approximated as
a discrete time Markov process with time discretized typically to
the channel coherence time \cite{goldsmith-book}. The set of all
possible fading gains (equivalently, SNR or BER levels) is modeled
as the states of the MC, and the channel variations are modeled as
the transitions between these states that occur at each interval
with certain probabilities. We used a two-state (Gilbert-Elliot)
model, characterized by the following parameters specified in
\cite{wireless-model}: (i) the bit-error-rate (BER) of each state,
which depends on the channel signal-to-noise (SNR) level (ii) the
state duration and transition probabilities, which depend on the
channel coherence time, and in turn on the speed of mobiles and the
channel frequency. We assume nomadic or pedestrian wireless clients
with 3 km/h speed, and 2.4 GHz channel frequency, which results in
an average coherence time of 21 ms. This duration ensures that the
channel remains static during a packet transmission. In our
experiments, we consider different channel quality levels assigning
average SNR levels from the set $\{3, 5, 7, 9\}$ dB, resulting in
effective packet loss rates from $1\%$ to $35\%$.

\subsubsection{Video Sequences}
As our test sequence, we used standard sequences:
\emph{Carphone}, \emph{Foreman}, \emph{Mother \& Daughter}.
These were QCIF sequences encoded using the JM 8.6 version of
the H.264/AVC codec \cite{codec, software}. The group of
pictures consisted of I and P frames, one I every 10 frames.
All encoded sequences had data rates of 70 kbps each and frame rate
of 30 fps. Each frame consists of at least one slice. Each slice was
packetized into an independent NAL (network abstraction layer)
unit, of size 250B. There are two reasons for this choice of packet
size. First, due to the time varying nature of wireless channels,
it is preferred to have short packet transmissions to avoid the
channel variation during a packet's transmission. Second,
using a fixed size (on the average) simplifies the network coding
(bit-wise XOR) operations requiring small or no padding of the packets.

The average peak-signal-to-noise (PSNR) ratio for the encoded
sequences, Carphone, Foreman and Mother\& Daughter, were $29.95dB$,
$28.70dB$ and $40.74dB$ respectively; these PSNR values, of the
encoded sequences before transmission, are denoted as ``No Error''
in the first row of Table \ref{table:avgPSNR} and correspond to the
top lines of Fig. \ref{perf_fig1} . We repeated and concatenated the
standard sequences to create longer test sequences of duration 30sec
each.

We simulated packet loss by erasing the corresponding NAL units from
the RTP stream, according to the packet traces produced by the
network simulation. At the receiver side, we decoded the remaining
RTP stream with standard error concealment enabled. When an entire
frame was lost, we used copy-concealment.

\subsubsection{Algorithms under Comparison}
We compare our algorithms, NCV and NCVD, against two baseline
algorithms for packet scheduling: no Network Coding (noNC) and
Network Coding for Throughput (NCT), which are described next.

{\bf No Network Coding (noNC):} This is a FIFO Tx queue
without network coding. Consider again Example \ref{ex1}
and Fig. \ref{fig:ncv}: node $I$ stores packets for all three streams
destined to nodes $A,B,C$. In every time slot, $I$ transmits the
first packet from the head of the queue. It may require several
consecutive retransmissions until the head-of-line packet is
successfully transmitted; in 802.11, there is an upper limit in the
maximum number of allowed retransmissions. In order to conduct a
fair comparison with our schemes, we slightly modify (improve) this
scenario by using the same scheme as described in section
\ref{sec:system}.\footnote{\label{footnote:scheme}The summary of the
scheme is as follows. After transmission, a packet is marked as
\emph{inactive} and is not transmitted as primary for a time
duration of a mean RTT; during that period, other packets are
transmitted from the FIFO as primary, thus better utilizing the
channel. After an RTT, if an ACK is still not received,
the packet is marked as \emph{active} and considered again for
transmission. Packets whose playout deadlines have expired are
removed from the Tx queue.}

{\bf Network Coding for Throughput (NCT):} This is an improved
version of the algorithm proposed in \cite{xor}. The packet
transmission mechanism is the same as in the {\em noNC} scheme, but
network coding is used to maximize throughput, as follows. The
packet at the head of the Tx queue is selected as a primary packet;
side packets are chosen to be XOR-ed together with the primary
packet so as to construct a network code that is useful to the
maximum number of receivers possible at this time slot.

There are two improvements in NCT compared to the coding algorithm
in \cite{xor} that allow NCT to achieve even higher throughput than
\cite{xor}. First, NCT follows the same ACK and retransmission
mechanism described in section \ref{sec:system} and repeated in
Footnote \ref{footnote:scheme}: packets with pending acknowledgments
are marked as inactive for one RTT, while the channel is used to
transmit other packets as primary. In \cite{xor} and in general MAC
retransmissions, a packet stays at the head of the queue blocking
other packets, until it goes through successfully or it exceeds the
maximum number of retransmissions. Head-of-line blocking avoids
reordering, which may be an issue for TCP traffic. However, in the
context of video streaming, the playout buffer can handle
reordering. Another difference is that NCT uses an improved version
of the coding procedure in \cite{xor}: NCT considers all possible
subsets of the candidate side packets thus maximizing the number of
receivers that can decode; while \cite{xor} considers side packets
in a sequential order, thus sacrificing some throughput for reduced
complexity. Therefore, we use NCT  as our baseline for the maximum
achievable throughput per transmission using network coding.

The main difference between NCV/NCVD and NCT is that our schemes
select side packets to maximize video quality while NCT maximizes
throughput. A secondary difference, is that we consider all packets
in a queue as candidates for side packets, while NCT, consistently
with \cite{xor}, considers only the earliest packet per
flow.

\subsection{Simulation Results}
In this section, we present simulation results that compare our
schemes to the baselines and demonstrate that NCV and NCVD can
improve video quality and application-level throughout, without
compromising MAC-level throughput. We report simulation results for
the single-hop scenario of Fig. \ref{fig:ncv}, when node $I$ streams
sequences \emph{Carphone}, \emph{Foreman}, and \emph{Mother and
Daughter} streamed to clients $A,B,C$, respectively.

\subsubsection{Video Quality Improvements}
\begin{figure*}
\begin{center}
\begin{tabular}{ccc}
\subfigure[Carphone]{{\includegraphics[width=5cm]{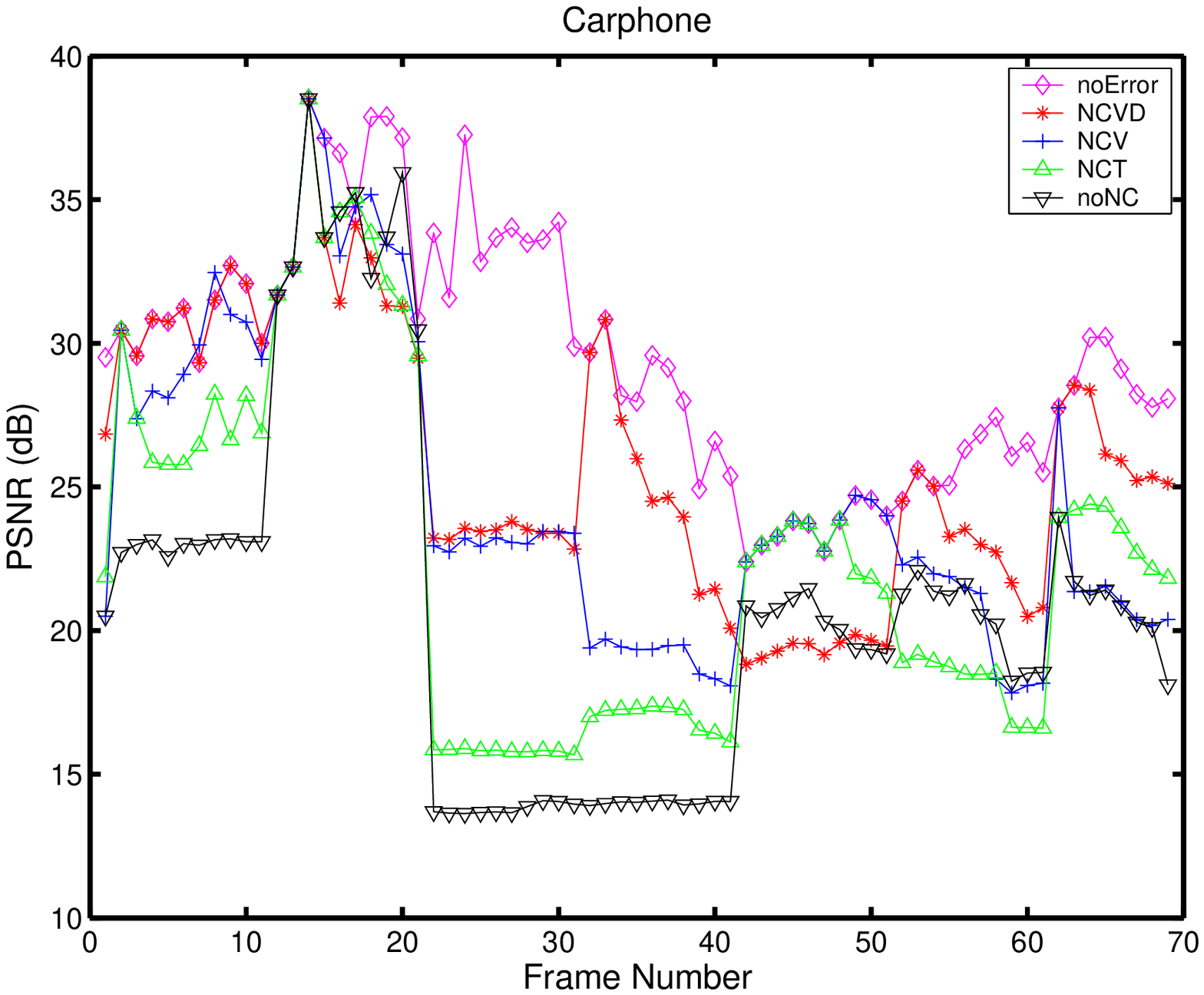}}}
\subfigure[Foreman]{{\includegraphics[width=5cm]{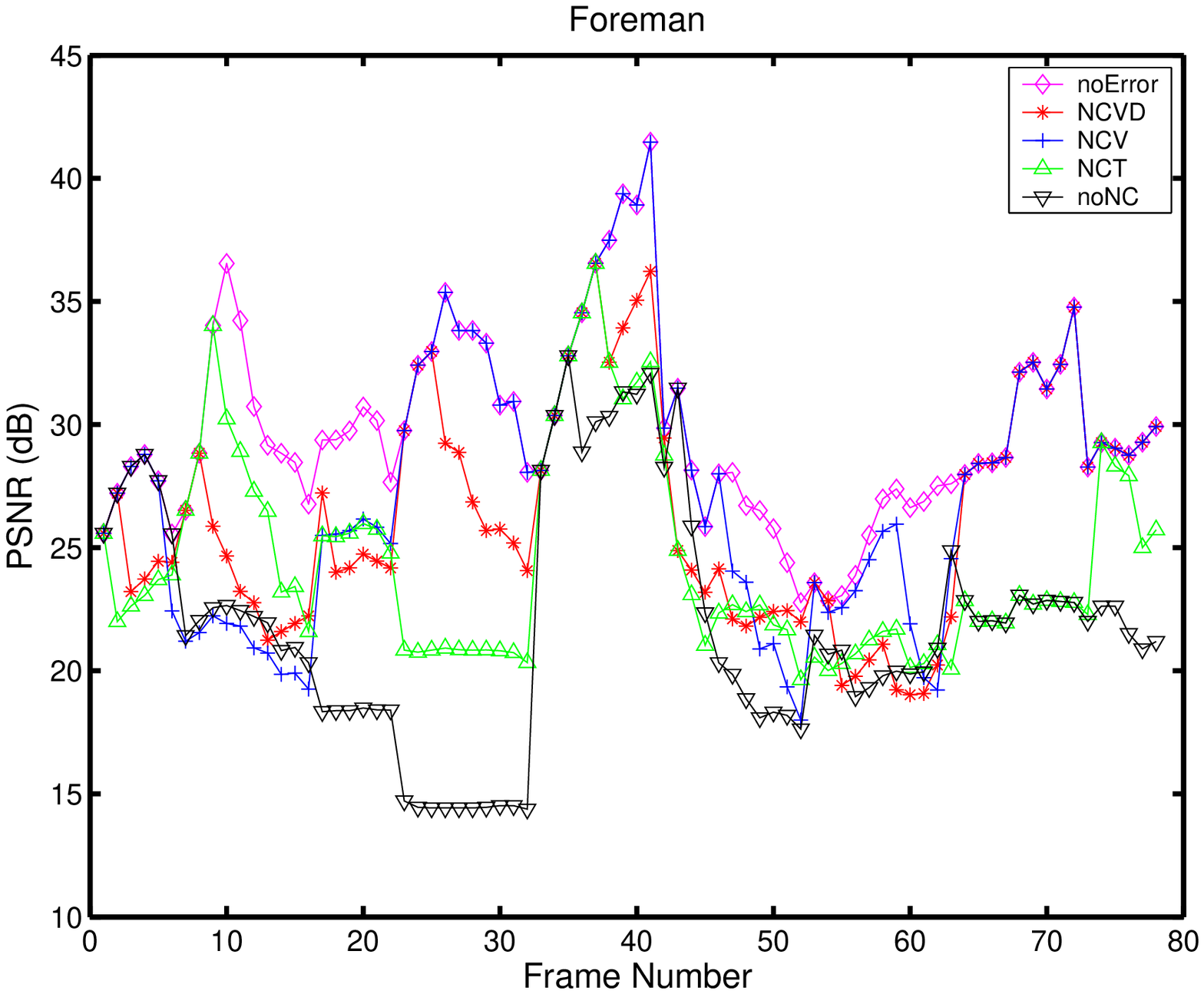}}}
\subfigure[Mother \&
Daughter]{{\includegraphics[width=5cm]{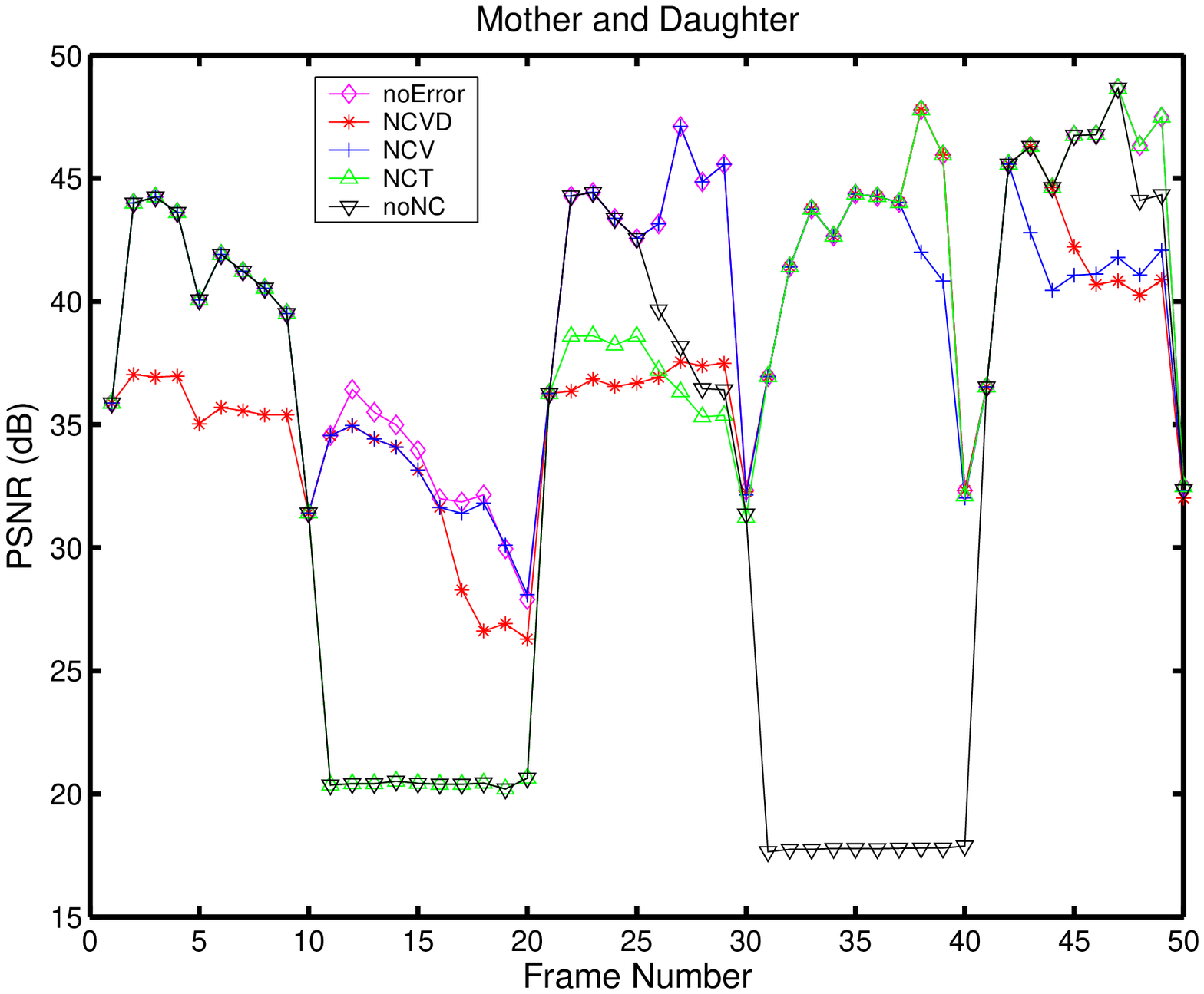}}}
\end{tabular}
\end{center}
\caption{PSNR per frame for (parts of) the test sequences, for
channel Model I with packet loss $9.4\%$ and delay budget 100ms.
Four schemes (noNC, NCT, NCV, NVCD) are compared. ``NoError'' refers
to the encoded sequences without any loss. The average PSNR values
(averaged over the entire sequence) are summarized in Table
\ref{table:avgPSNR}).} \label{perf_fig1}
\end{figure*}

Fig.~\ref{perf_fig1} shows the video quality experienced by the
clients (PSNR over frame number for parts of the sequences) for the
four algorithms under comparison, namely noNC, NCT, NCV, NCVD, as
well as for the encoded sequences before transmission (noError). The
simulation is performed for wireless channel \emph{Model I} at
packet loss rate of $9.4\%$ with 100 ms delay budget; for
comparison, the same wireless channel trace is used as input to all
4 algorithms. As expected, there are time periods, during which the
channel is bad, the quality degrades for all algorithms. However,
the degradation for NCV and NCVD is much less than for NCT and noNC,
because NCV and NCVD select network codes to protect and deliver the
most important packets on time, thus improving the video quality; in
contrast,  NCT and noNC treat all packets similarly.

The average PSNR for each sequence and algorithm is summarized in
Table \ref{table:avgPSNR}. As expected, the noNC scheme performs
poorly. NCT improves over noNC because it delivers more packets per
time slot. NCV improves over NCT because it chooses the most
important video packets; although the number of packets does not
increase over NCT, their quality does. NCVD further improves over
NCV because it considers more candidate codes and opportunities.
These numbers are compared to the original encoded sequence
(NoError).

\begin{table}
\caption{\label{table:avgPSNR}Average PSNR for the scenario of
Fig.\ref{perf_fig1} ($70kbps$ video rate, channel Model I with
$9.4\%$ loss, $100ms$ playout deadline)}
\begin{center}
\begin{tabular}{|c||c|c|c|}
\hline
avg PSNR (dB) & Carphone & Foreman & Mother\&Daughter\\
\hline
No Error & 29.95 &28.70& 40.74 \\
NCVD & 26.32 &26.08 &32.87\\
NCV &23.99 &25.03 &32.62 \\
NCT &22.40 &22.76 &30.81 \\
noNC &22.08 &21.59 &26.92 \\
 \hline
\end{tabular}
\end{center}
\vspace{-15pt}
\end{table}

\begin{figure}
\begin{center}
{\includegraphics[width=5.3cm]{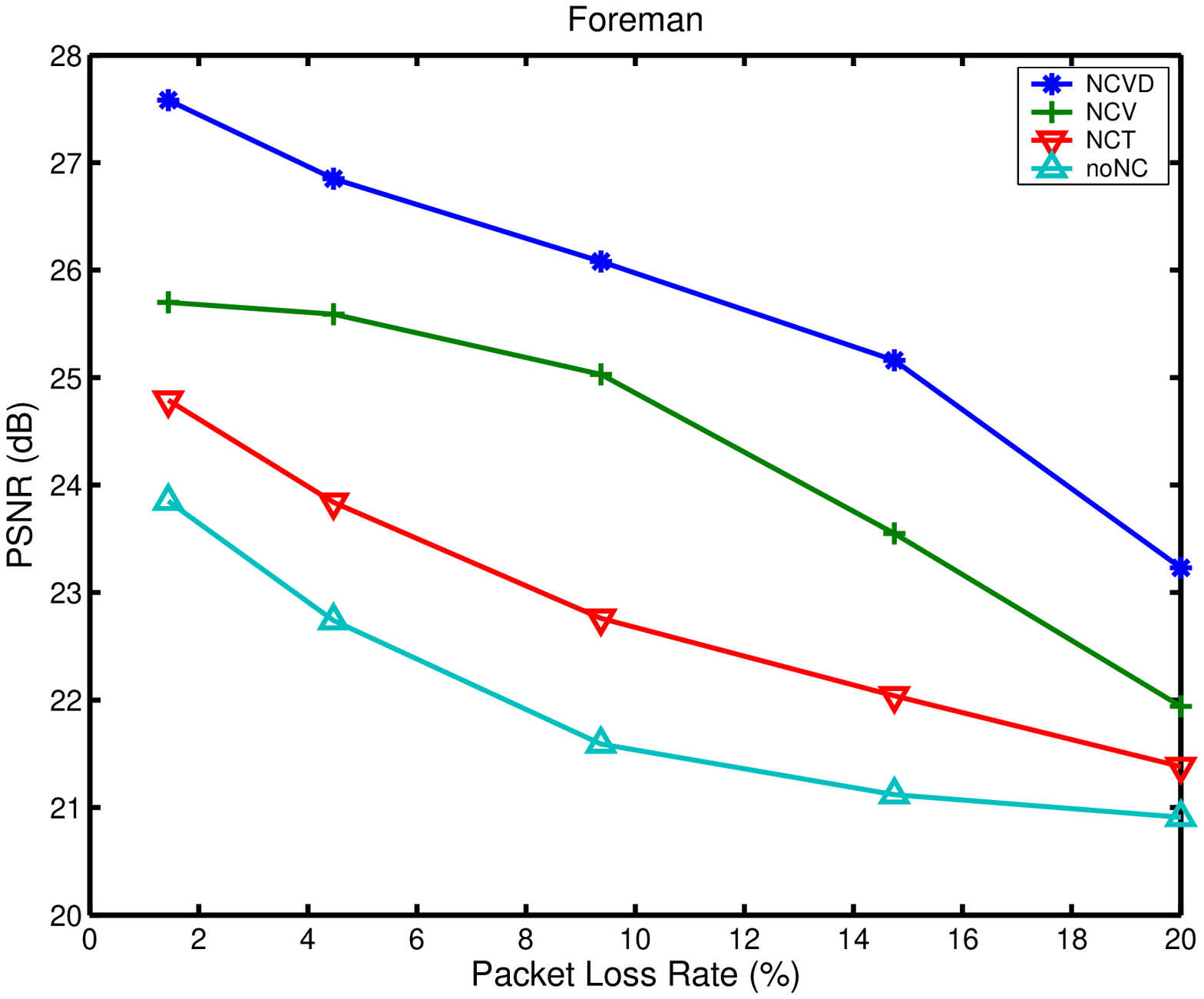}}\\
{\includegraphics[width=5.3cm]{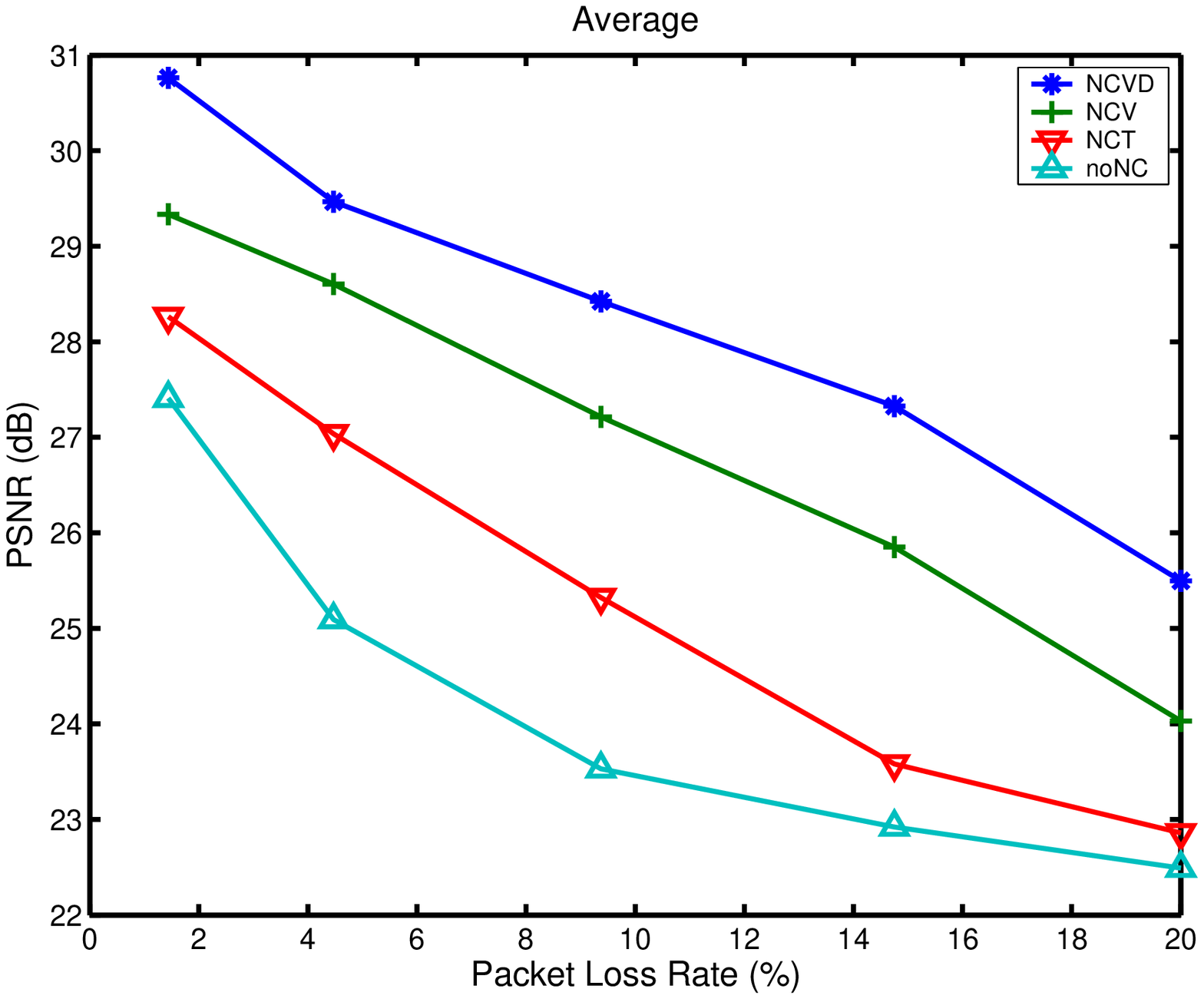}}\\
\end{center}
\caption{Video quality observed over wireless channel \emph{Model
I}. Average PSNR for \emph{Foreman}, and Average PSNR, averaged
across all three sequences. } \label{perf_fig2}
\end{figure}

Fig.~\ref{perf_fig2} focuses on the client that receives the
sequence \emph{Foreman}. The same scenario as in Fig.~\ref{perf_fig1}
is considered, but with loss rates varying from 1\% to 20\%.
Fig.~\ref{perf_fig2} shows the average PSNR for each value of packet
loss rate and for each algorithm. Clearly, NCV and NCVD outperform
NCT (by $2.5-3.5$dB) and noNC (up to $3.5-5$ dB) for all packet loss
rates. Another observation (from this and other figures omitted for
lack of space) is that the PSNR gain of NCV and NCVD is larger for
medium than for very low and very high packet loss rates. For low
loss rates, most packets are transmitted successfully, while for
high loss rates most packets are lost. In both cases, the
number of network coding opportunities decrease. However, even then,
the proposed algorithms still achieve a considerable PSNR
improvement. The upper part of Fig.~\ref{perf_fig2} shows the video
quality for the Foreman sequence, and the lower part shows the PSNR
averaged across all sequences.

\begin{figure}
\begin{center}
{\includegraphics[width=5.3cm]{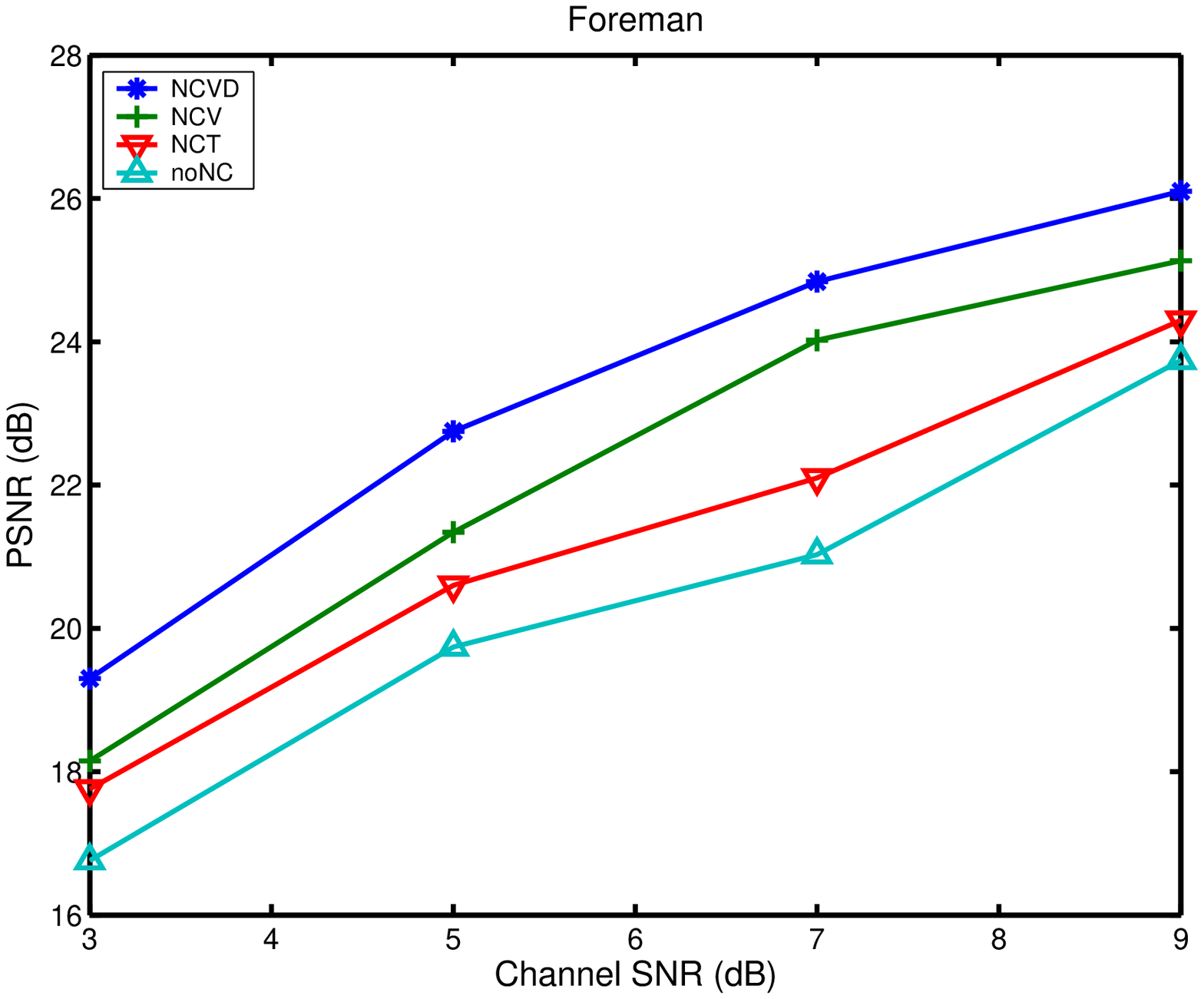}}\\
{\includegraphics[width=5.3cm]{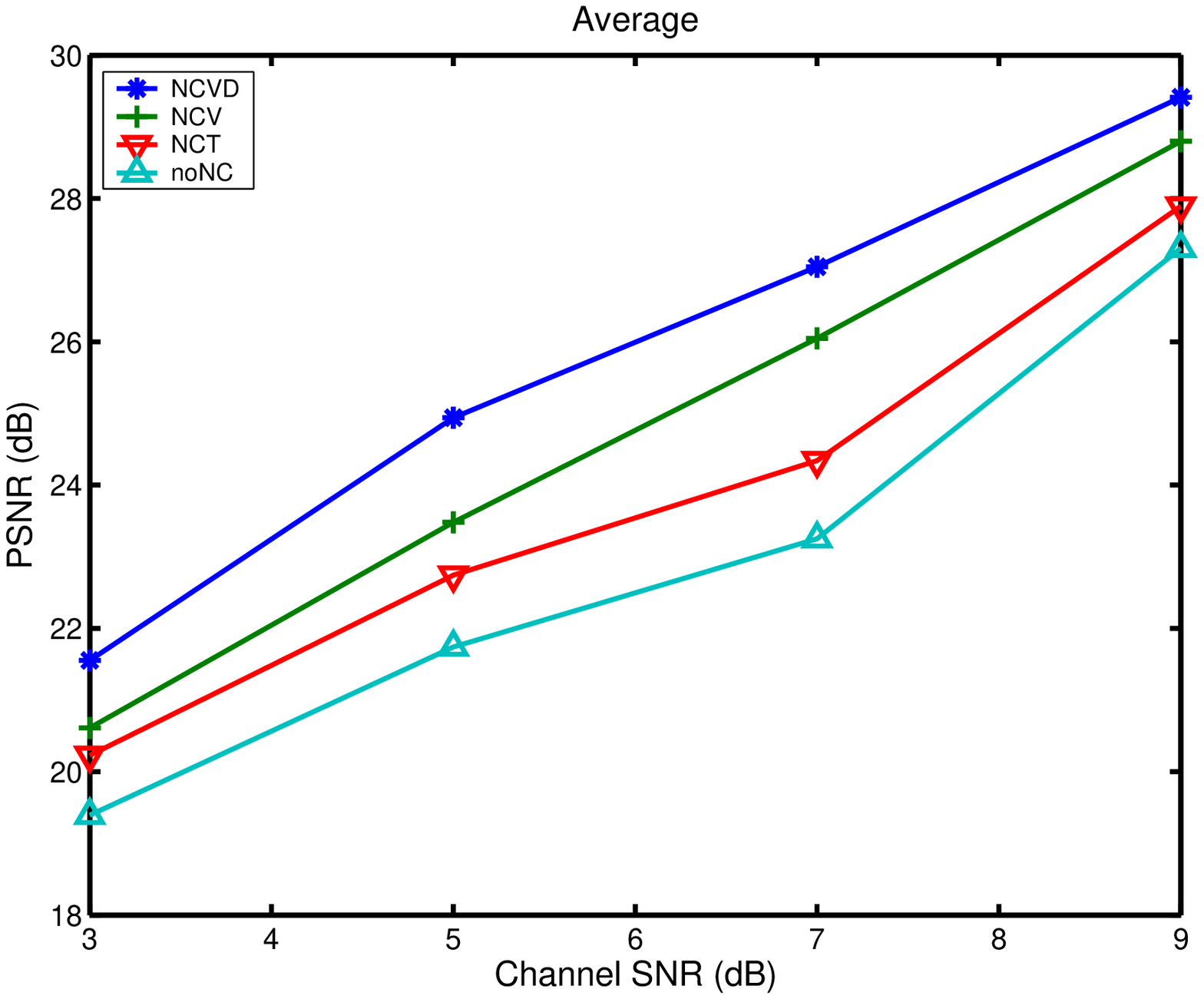}}\\
\end{center}
\caption{Video quality observed over wireless channel \emph{Model
II}. Average PSNR for \emph{Foreman}, Average PSNR, averaged
across all three sequences.} \label{perf_fig3}
\end{figure}

Fig.~\ref{perf_fig3} evaluates the same scenario as in
Fig.\ref{perf_fig2}, but for the second wireless channel \emph{Model
II}. This is the two-state model, parameterized by the SNR levels
(resulting in loss rates from 1\% to 35\%). The compared algorithms
are ranked similar to the previous case and the PSNR improvements
from NCV and NCVD are still high. However, there is less improvement
compared to channel \emph{Model I}. The difference can be explained
by the network coding and code selection opportunities. In
\emph{Model I}, consecutive transmissions to the same client
experience independent loss, and links to different clients are
independent from each other. This independence
results in all virtual buffers having roughly the same number of
packets, over short time periods. However, in \emph{Model II}, the
channel alternates between a good and a bad state; while in a bad
state, a client is more likely to experience consecutive losses,
while clients in a good state are more likely to receive consecutive
successful packets. This results in an unbalanced number of packets
in the virtual buffers. Since network code construction directly
depends on the number of packets in virtual buffers, there are less
network coding opportunities, hence less network code selection
possibilities. Even when there are less opportunities, we still
observe significant quality improvement: NCV and NCVD improves up to
2 and 3dB over NCT and up to 3 and 4dB over noNC, respectively.


\begin{figure}
\begin{center}
{\includegraphics[width=6cm]{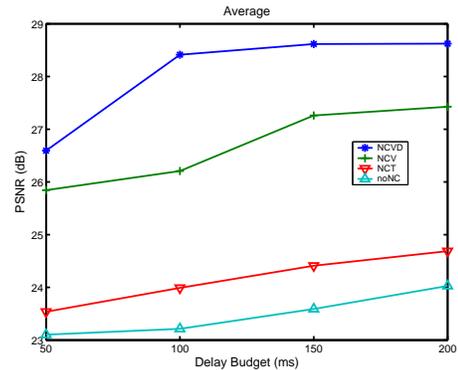}}\\
\end{center}
\caption{PSNR values (averaged over each sequence and across
sequences) for different delay budgets. Wireless {\em Model I} is
considered.\label{perf_fig4}}
\end{figure}

In the previous scenarios, we considered a delay budget of 100ms. We
now consider channel Model I at $9.4\%$ packet loss rate, with delay
values ranging from $50$ to $200$ms. Fig.~\ref{perf_fig4} depicts
the PSNR values for each scenario (first averaged over each sequence
and then across the three sequences, to summarize the overall
improvement). The figure shows that both NCV and NCVD improves video
quality for the entire range of delay values. The improvement is
smaller for a tight delay budget, because a tight delay constraint
limits the number of retransmissions and the lifetime of packets
both at the Tx queue and the virtual buffers, thus decreasing
network coding and selection opportunities. However, even with tight
delay constraints, there is significant video quality improvement
from NCV and NCVD compared to NCT and noNC.

\subsubsection{Throughput Improvements}
\begin{figure*}[t!]
\begin{center}
\begin{tabular}{ccc}
\subfigure[Application-level throughput over channel {\em Model
I}]{{{\includegraphics[width=5cm]{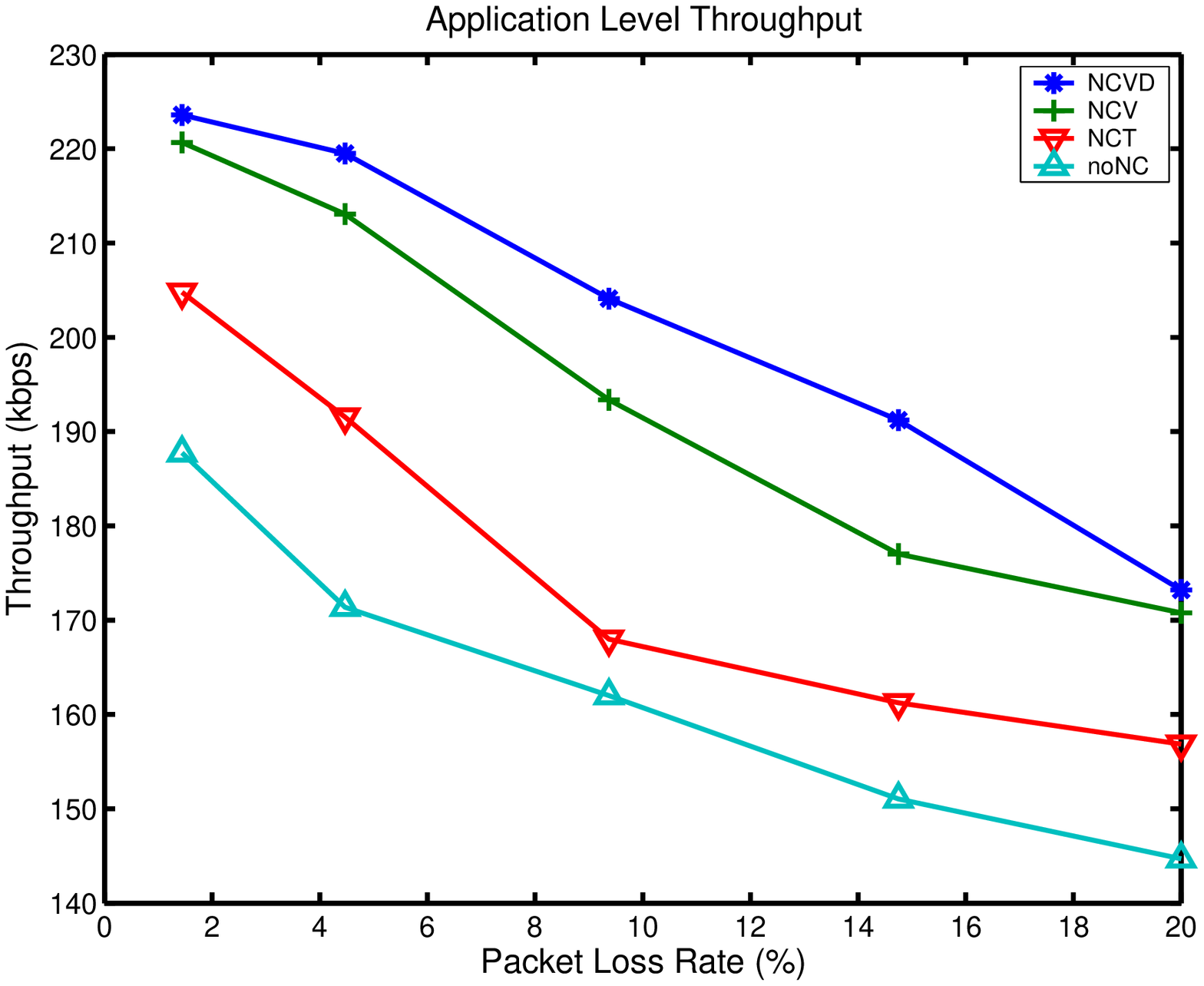}}}}
\subfigure[Application-level throughput over channel {\em Model
II}]{{{\includegraphics[width=5cm]{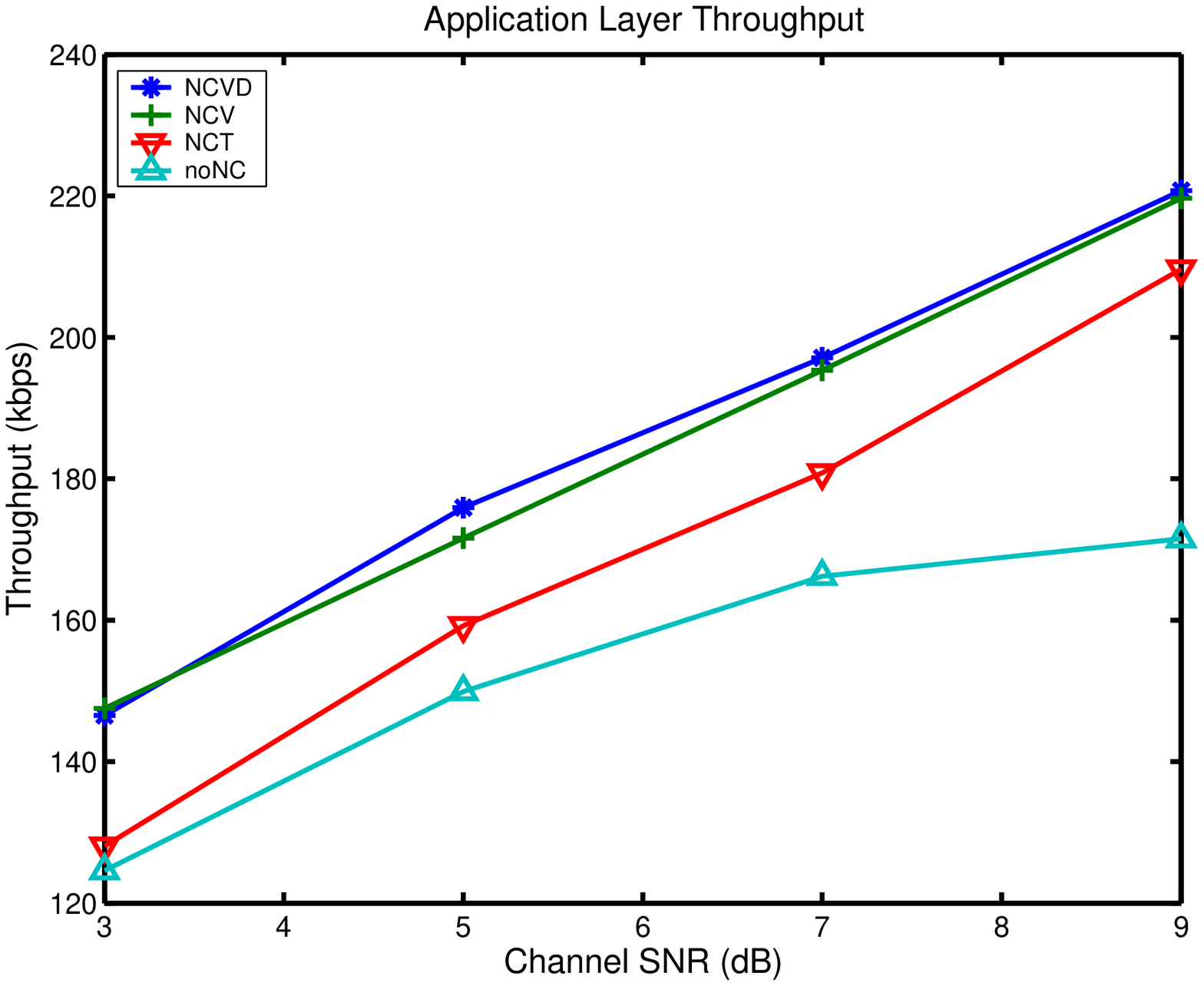}}}}
\subfigure[MAC-level throughput over channel {\em Model
I}]{{{\includegraphics[width=5cm]{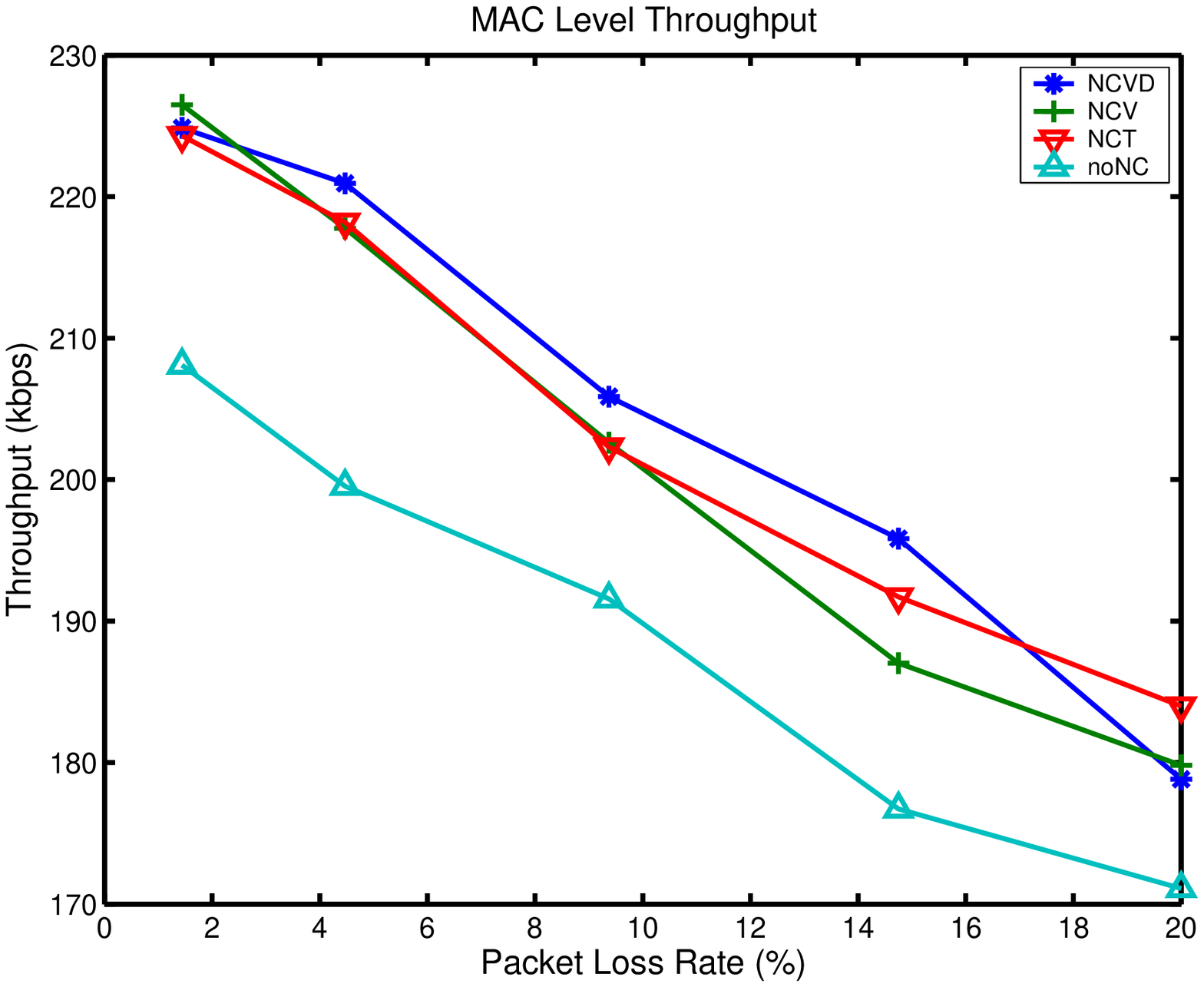}}}}
\end{tabular}
\end{center}
\caption{Total throughput (added over all three streams) achieved by
the four algorithms. \label{perf_fig5}}
\end{figure*}

The video-aware schemes improve video quality because they
explicitly take it into account in the code selection. In this
section, we show that, our schemes also significantly
improve application-level throughput.

{\em Application throughput.} Fig.~\ref{perf_fig5}(a) and
(b) show the total throughput as seen by the
application-layer (i.e. NAL units per sec) added over all
clients. The figure clearly shows that NCV and NCVD achieve higher
throughput as compared to NCT and noNC. The main reason
is that NCV and NCVD do not select codes consisting of
packets whose deadlines are within one transmission time, while  NCT
and noNC transmit all packets. Late packets does not contribute
to application-level throughput, because those
packets are discarded at the client even if they are received
successfully. NCV and NCVD transmit more useful packets to the
client, even though the number of transmitted packets may be
comparable to NCT. Application-level throughput is the most relevant
notion of throughput in our context, as in \cite{xor}.

{\em MAC Throughput.} For completeness, we also show that the MAC-layer
throughput of NCV and NCVD is very close to that of NCT, which is
specifically designed for maximum throughput.
Fig.~\ref{perf_fig5}(c) shows the MAC-level throughput for the four
schemes. As expected, NCT, NCV and NCVD all achieve higher MAC-level
throughput than noNC; this is because all three schemes use network
coding across streams, which increases the information content per
transmission. The second and more interesting observation is that
NCV and NCT achieve almost the same MAC level throughput, for loss
rates up to $9.4\%$, and NCT achieves slightly higher throughput for
larger loss rates. The reason is that NCV and NCT use the same
network code construction schemes but different network code
selection mechanisms; therefore, the amount of data delivered at the
MAC layer is almost the same for low packet loss rates. For higher
packet loss rates, there are more packets with urgent deadline in
the Tx queue. NCT transmits all possible side packets even if the
their deadline is within one transmission time. NCV selects both
primary and side packets considering the deadline and importance;
therefore, NCV may purposely {\em not} choose the code with the
largest number of packets,  if one or more packets in this code are
going to be useless at the receiver. Instead, NCV
sends the most useful and effective packets to the client.
In summary, NCV achieves the same MAC-level throughput with NCT for
loss rates up to 9.4\%, and slightly less for higher loss rates.
However, and more importantly, NCV always achieves higher
application-level throughput than NCT.

{\em NCVD.} A third observation from Fig.\ref{perf_fig5}(c) is that
NCVD achieves not only the highest application-level throughput but
also the highest MAC throughput for most loss rates. This is
explained by the fact that NCVD looks into the entire queue and has
more options to choose from, both in terms of video quality and in
terms of absolute number of packets. For very high loss rates (above
$20\%$) NCT achieves higher MAC throughput because NCVD prefers to
optimize the code selection for the application level.

{\em Small Queue Sizes.} We finally looked at the buffer occupancy
at the transmitter and at the clients and observed that they
were really small, in the order of 5-10 packets, for the simulated
scenarios and delay budgets considered.
Furthermore, we observed that NCV and NCVD further reduces
the queue size, compared to NCT and noNC. This is intuitively expected: NCV
and NCVD deliver more packets successfully to the client, thus
there are less packets waiting for transmission in Tx.
Maintaining short buffers has several positive implications. First,
short queues are good for bottlenecks shared with TCP. Second,
having a small Tx queue allows to consider all packets in the queue
for network codes and enjoy the performance gains of NCVD without
increase in complexity, even for $d=\infty$. Finally, having small
virtual buffers means that there are only a
few overheard packets to consider in the construction of the network
codes, which significantly decreases the complexity.

\section{\label{sec:discussion}Discussion}

\subsection{Video Quality and Throughput}
Our metric is the improvement in video
quality for every single node, in Eq.~(\ref{eqNo2}), added over all
nodes, in Eq.~(\ref{eqNo1}). Although, throughput is not explicitly
mentioned, it is implicitly captured. E.g.
if we omit the distortion term $(1-e_{l}^{k})\Delta_{l}^{k}$ from
Eq.~(\ref{eqNo2}), Eq.~(\ref{eqNo1}) simply becomes
$I_k^{i}=\sum_{m=1}^{N} \sum_{l=1}^{L_k}g_{l}^{k}d_{l}^{k}$, which counts the
absolute number of packets delivered in a single transmission, i.e.
throughput, similarly to NCT and to \cite{xor}. Therefore, our
metric successfully captures both the number and the quality of
packets in a code. Furthermore, our schemes choose
``better-quality'' packets and improve the application-level
throughput, as already shown in the performance evaluation.

\subsection{\label{sec:complexity}Complexity Analysis}
NCV (in Alg.~\ref{algo:NCV}) constructs candidate network codes and
selects the one that maximizes video quality improvement. The main
complexity comes from considering {\em all} possible candidate
codes. Selecting side packets among all possible subsets
of overheard packets at the target node, is clearly exponential in the
size of the virtual buffer. However, the complexity of NCV is no worse than the complexity of
NCT: they both consider all possible codes but they evaluate them
using a different metric. Another important observation is that
real-time delay requirements significantly reduce the number of
packets in the virtual buffers and therefore the complexity; e.g. in
our simulations, a delay budget of 100ms resulted in at most $5$
packets in the virtual buffers. Thus, the brute-force approach is feasible for
real-time applications.

For a larger delay budget, approximation algorithms for NCV and NCT can be developed as
follows. Recall that in each time slot, first we pick the
primary packet, then we look at all overheard packets at the target
client as candidate side packets. Conflicts between side packets can
be represented using a graph, whose vertices represent the overheard
packets and edges represent their conflict. Each vertex has a weight
corresponding to the expected improvement from including this packet
into the code. Two vertices are connected through an edge if the
corresponding packets cannot be used together in the same code.
The problem of selecting side packets to maximize quality
improvement is then reduced to selecting vertices that are not
connected to maximize the total weight. This problem is known
as the maximum weight independent set problem and
is the complement of the vertex cover problem. Although these
problems are NP-complete, they are also well-studied,
and approximation algorithms can be found in the literature,
\cite{app-algorithms}.

Finally, NCVD (Alg.~\ref{algo:NCVD}) runs NCV for each packet
(considered as primary) in the Tx queue and selects the best overall
code. The NCVD complexity is linear in the number of packets in the
Tx queue, which is also small for real-time applications. The
dominant part is still due to the NCV part.

\subsection{Ongoing and Future Work}
We are currently extending our approach to several directions. We
are experimenting with additional topologies/scenarios, beyond the
last-hop one-directional scenario discussed in this paper.
We are also exploring extensions of our framework to capture (i) the
dependency among video packets (ii) the benefit of
overheard packets and (iii) code selection across multiple time
slots.

\section{\label{sec:conclusion}Conclusion}
In this paper, we proposed a novel approach to opportunistic video
coding for video streaming over wireless networks that takes into
account the importance of video packets in network code selection.
Simulation results show that the proposed schemes improve video
quality up to $3-5$dB compared to baseline schemes. Furthermore,
they significantly improve the application-level throughput and
achieve the same or similar levels of MAC throughput.

\end{document}